# LIEB-WU SOLUTION, GUTZWILLER-WAVE-FUNCTION, AND GUTZWILLER-ANSATZ APPROXIMATIONS, WITH ADJUSTABLE SINGLE-PARTICLE WAVE FUNCTION FOR THE HUBBARD CHAIN

Jan Kurzyk[1*], Jozef Spałek[2†], Włodzimierz Wójcik[1*]
[1]*Institute of Physics, Politechnika Krakowska, ulica Podchorążych 1, 30-084 Kraków, Poland*
[2]*Marian Smoluchowski Institute of Physics, Uniwersytet Jagielloński, ulica Reymonta 4, 30-059 Kraków, Poland*

The optimized single-particle wave functions contained in the parameters of the Hubbard model ($t$ and $U$) were determined for an infinite atomic chain. In effect, the electronic properties of the chain as a function of interatomic distance $R$ were obtained and compared for the Lieb – Wu exact solution (LW), the Gutzwiller-Wave-Function approximation (GWF), and the Gutzwiller-ansatz case (GA). The ground state energy and other characteristics for the infinite chain were also compared with those obtained earlier for a nanoscopic chain within the **E**xact **D**iagonalization combined with an *Ab Initio* adjustment of the single-particle wave functions in the correlated state (*EDABI method*). For the sake of completeness, we briefly characterize also each of the solutions. Our approach completes the Lieb-Wu solution, as it provides the system electronic properties evolution as a function of physically controllable parameter – the interatomic distance.

---

[*] Two of authors (W.W. and J.K.) would like to dedicate this work to Prof. J.Spałek on his 60th Birthday.
[†] coresponding author; ufspałek@if.uj.edu.pl; url: http://th-www.if.uj.edu.pl/ztms/jspalek_een.htm

## 1. Introduction

The question of a proper description of electronic states in correlated electron systems is regarded as one of the most important problems in condensed matter physics. This is because the single-particle approach such local density approximation (LDA) is usually insufficient and must be supplemented by many-body corrections. In effect, the methods such as LAD+U, LDA+Dynamic Mean-Field Theory (LDA+DMFT) have been developed, the range of validity of which has not been fully tested.

The difficulties are caused by the fact that in correlated systems the single-particle part of the total energy (the band or kinetic energy) is usually comparable or even smaller than the Coulomb interaction part [1]. Hence the electron – electron interaction cannot be regarded as a smaller contribution to the total energy of correlated systems. Having in mind the difficulty we have devised a different approach [2]. Namely, we take the exact solution of the parameterized model in the second quantization representation, in which the parameters contain in a functional manner the single-particle wave functions, and determine those functions *a posteriori* by treating the ground state energy as a functional of those wave functions. As a result, we obtain the *renormalized* (self-adjusted) wave equation (SWE) for those wave functions, which include implicitly the effect of correlations. This method combines in a natural manner the $2^{nd}$ and $1^{st}$ quantization schemes and, in principle, the only approximation made is the limitation of the basis, which is explicitly contained in the definition of the model considered [3].

This approach has been applied so far only to limited number of situations. We applied it to the nanoscopic systems and have addressed the question concerning the "Mott physics" as applied to a nanoscopic scale [3,4]. Here we apply the same scheme to the one-dimensional Hubbard model, for which the exact solution of Lieb-Wu [5] is available. We compare this solution with both the Gutzwiller-ansatz (GA) [6] and the Gutzwiller-wave-function (GWF) [7] approximations. The last two approximation schemes, particularly the Gutzwiller-ansatz solution, can be applied to the system of higher dimensions, where it serves as a starting point to a more sophisticated, albeit approximate, analysis of electronic properties [8]. The variational solution of SWE discussed here demonstrates the feasibility of the method as applied to the extended systems and, in fact, completes the solution of parameterized models by providing the system evolution as a function of physically controllable parameter – the lattice parameter.

The structure of the paper is as follows. In Sec. 2 we define the problem, as well as characterize briefly the Lieb-Wu (LW), Gutzwiller-ansatz (GA), and Gutzwiller-wave-function (GWF) solutions. In Sec. 3 we discuss the ground state properties of the Hubbard chain, as well as determine the correlation-induced wave function spatial-extension change. We also compare there our variational results for intensive quantities with those obtained earlier for nanoscopic systems when the same boundary conditions (periodic) are taken in both situations. Sec. 4 contains conclusions.

## 2. Starting Hamiltonian

We consider the extended Hubbard model for correlated narrow-band electrons in a linear chain (*the extended Hubbard chain*). This model is represented by the Hubbard Hamiltonian of the form

$$H = \varepsilon_a \sum_i n_i + \sum_{i\sigma} t_{ij}(a_{i\sigma}^+ a_{i+1\sigma} + h.c.)$$
$$+ U \sum_i n_{i\uparrow} n_{i\downarrow} + \sum_{i<j} K_{ij} n_i n_j + \sum_{i<j} V_{\text{ion}}(\mathbf{R}_i - \mathbf{R}_j), \qquad (2.1)$$

where $\varepsilon_a$ is the atomic energy of electron, $t_{ij}$ – hopping integral for the atomic sites $i$ and $j$, $U$ is the intratomic part of the Coulomb energy, $K_{ij}$ – is the Coulomb interaction for electrons located on sites $i \neq j$, and

$$V_{\text{ion}} \stackrel{a.u.}{=} \frac{2}{|\mathbf{R}_i - \mathbf{R}_j|} = \frac{2}{R_{ij}}, \qquad (2.2)$$

is the ion-ion Coulomb repulsion (in atomic units a.u.) between ions located at the distance $R_{ij}$. We can rewrite the intersite part of the electron-electron interaction in the form

$$\sum_{i<j} K_{ij} n_i n_j = \sum_{i<j} K_{ij}(n_i - 1)(n_j - 1) - \sum_{ij} K_{ij} + 2 N_e \frac{1}{N_a} \sum_{i<j} K_{ij}$$
$$= \sum_{i<j} K_{ij} \delta n_i \delta n_j + N_e \frac{1}{N} \sum_{i<j} K_{ij} + (N_e - N_a) \frac{1}{N_a} \sum_{i<j} K_{ij}, \qquad (2.3)$$

where $N_e$ is the total number of electrons in the system, $N_a$ is the number of atoms (ions), and $\delta n_i = n_i - 1$. In the considered here *Mott insulating state* we have on average $\langle \delta n_i \rangle = 0$, achieved for $N_e = N_a$ (the narrow band is half-filled, $n = 1$) and therefore

$$\sum_{i<j} K_{ij} n_i n_j = \sum_{i<j} K_{ij}. \qquad (2.4)$$

In effect, Hamiltonian (2.1) reduces to the effective Hubbard-model form

$$H = \varepsilon_a^{\text{eff}} \sum_i n_i + \sum_{i\sigma} t_{ij}(a_{i\sigma}^+ a_{i+1\sigma} + h.c.) + U \sum_i n_{i\uparrow} n_{i\downarrow}, \qquad (2.5)$$

where

$$\varepsilon_a^{\text{eff}} \equiv \varepsilon_a + \frac{1}{N} \sum_{i<j} \left( K_{ij} + 2/R_{ij} \right) \qquad (2.6)$$

is the effective atomic energy. Note that such a redefinition of $\varepsilon_a$ will provide a correct value of atomic energy in the limit of large interatomic distance ($R_{ij} \to \infty$), i.e. in the atomic limit, since then we have to include electron-ion, electron-electron, and ion-ion interactions to reach the limit of *neutral* atoms. This redefinition is necessary as we are going to discuss the properties of the systems as a function of interatomic distance (not only as a function of $U/t$). For the sake of completeness, we write down explicitly the expressions for the parameters

$$t_{ij} = \langle w_i | H_1 | w_j \rangle = \int d^3 r w_i^*(\mathbf{r}) H_1(\mathbf{r}) w_j(\mathbf{r}),$$

$$U = \langle w_i^2 | V | w_i^2 \rangle \stackrel{a.u.}{=} \int d^3 r d^3 r' | w_i(\mathbf{r}) |^2 \frac{2}{|\mathbf{r}-\mathbf{r}'|} | w_i(\mathbf{r}') |^2, \quad (2.7)$$

$$K_{ij} = \langle w_i w_j | V | w_i w_j \rangle \stackrel{a.u.}{=} \int d^3 r d^3 r' | w_i(\mathbf{r}) |^2 \frac{2}{|\mathbf{r}-\mathbf{r}'|} | w_j(\mathbf{r}') |^2.$$

In these definitions, $w_i(\mathbf{r})$ is the Wannier function centered on site $i$, and $H_1(\mathbf{r})$ is the Hamiltonian for a single particle in the medium. In that language, the starting atomic energy is

$$\varepsilon_a = \langle w_i | H_1 | w_i \rangle = t_{ii}. \quad (2.8)$$

Ground state energy per atom has the form

$$\frac{E_G}{N_a} \equiv \frac{1}{N_a} <H> = \varepsilon_a^{\text{eff}} + \frac{1}{N_a}\left(\sum_{i\sigma} t_{ij} <a_{i\sigma}^+ a_{i+1\sigma}> + U\sum <n_{i\uparrow} n_{i\downarrow}>\right). \quad (2.9)$$

Note that the $\varepsilon_a^{\text{eff}}$ energy appears explicitly here, since we will study the intersite-distance dependence of $E_G$. Under these circumstances, the atomic energy is not constant, as would be in the case if $U/t$ is the only parameter.

*2.1. Lieb – Wu solution*

Rigorous solution of the Hubbard chain was obtained some time ago by Lieb and Wu [5]. It is based on the so-called Bethe ansatz [9], in which the wave function of the system is constructed as follows. We postulate the many-particle wave function of the form

$$\Psi_Q(\mathbf{x},\mathbf{k}) = \sum_P [P,Q] \exp(i\sum_{j=1}^N k_{P_j} x_{Q_j}), \quad (2.10)$$

where $[P,Q]$ is the set of $N! \times N!$ ($N$ is the number of electrons, $N \equiv N_e$) of coefficients labeled with the permutations of $Q$ and $P$, which map the set of numbers $\{1,2,...,N\}$ onto the sets $\{Q_1, Q_2,..., Q_N\}$ and $\{P_1, P_2,..., P_N\}$, respectively. Permutation $Q$ describes the particle arrangement in the chain in the order

$$1 \leq x_{Q_1} \leq x_{Q_2} ... \leq x_{Q_N} \leq N_a, \quad (2.11)$$

whereas the permutation $P$ ascribes to the particles the numbers from the interval $-\pi < k \leq \pi$, with the ordering

$$k_1 < k_2 < ... < k_N. \quad (2.12)$$

As a result, we obtain the following expression for the ground state energy of electrons in the chain

$$\frac{E_G}{N_a} = \varepsilon_a^{\text{eff}} - 4|t|\int_0^\infty d\omega \frac{J_0(\omega) J_1(\omega)}{\omega[1+\exp(\omega U / 2|t|)]}, \quad (2.13)$$

where $J_0$ and $J_1$ are the Bessel functions zero-th and the first order, and $t$ is the value of the hopping element ($t < 0$) between the nearest neighbors in the chain.

*2.2. Gutzwiller-ansatz solution (GA)*

Gutzwiller [6] proposed variational solution of the Hubbard model. The trial function within that method is assumed in the form

$$\Psi = \prod_i \left[1 - (1-g)\hat{D}_i\right]\Psi_0, \qquad (2.14)$$

where $\Psi_0$ is the ground-state wave function for noninteracting fermions, $\hat{D}_i = n_{i\uparrow}n_{i\downarrow}$ is the number operator representing the number of double occupancies of electrons located on site $i$, which takes the eigenvalues 0 or 1, and $0 \leq g \leq 1$ is the variational parameter determined from the condition of $E_G$ minimum. For $g = 1$ the Gutzwiller wave function $\Psi$ reduces to $\Psi_0$ and corresponds to the $U = 0$ limit. As $U$ increases, or more precisely, $U/|t|$ increases, $g$ decreases and in the $U/|t| \to \infty$ reaches the asymptotic value $g = 0$. Note that in the latter limit $\Psi \neq 0$ is only when there are no double occupancies; this means that in the half-filled-band case we have reached the Mott-Hubbard-insulator limit. For the system with $n \equiv N/N_a \leq 1$, not only the value of $g$ decreases with the increasing $U/|t|$ ratio, but also the double occupancy decreases $\langle n_{i\uparrow}n_{i\downarrow}\rangle$, as it is energetically unfavorable. For $n > 1$ (i.e. for $N > N_a$), the number of unoccupied sites will decrease with the increasing $n$. In effect, the electron-hole symmetry can be used to discuss the situation with $n > 1$ in terms of that for $n < 1$.

The Gutzwiller wave function is simpler to work than that of (2.10). Nonetheless, it has been used without a further simplifications only for one dimensional systems [7] and for the hypercubic lattice of infinite dimension; $D \to \infty$ [10]. In general, we are forced to introduce the approximation that the electrons with the spin direction $\sigma = \uparrow$ move independently of those with $\sigma = \downarrow$. Under this circumstance, when considering motion of $\sigma = \uparrow$, those with $\sigma = \downarrow$ are regarded as infinitely heavy and vice versa. Obviously, due to the Coulomb (Hubbard) repulsive interaction, the two types of electrons try to avoid each other. So, the additional ansatz is applicable at best near *the Mott-Hubbard transition* point $(U/|t|)_c$. In that approximation, the whole approach reduces to the combinatorial problem of calculating the number of configurations. In this work we present the combined of single-particle wave function in the correlated state [2,3,11,12] with the exact [5] or approximate [6,10,13,14] of the correlations.

The original Gutzwiller approach is regarded sometimes as cumbersome. Therefore, in 1975 Ogawa, Kanda, and Matsubara [11] reformulated the approximation scheme. They have shown that the Gutzwiller approximation amounts to neglecting all spatial correlations but those between the nearest neighbors. Hence, the Gutzwiller solution acquired the name of *Gutzwiller-ansatz* (GA). In GA, the ground-state energy in $n = 1$ case is obtained explicitly in the following analytic form

$$\frac{E_G}{N} = \varepsilon_a^{eff} - |\bar{\varepsilon}|\left(1 - U/U_c\right)^2, \qquad (2.15)$$

where $\bar{\varepsilon}$ is the average kinetic (band) energy (per electron) for uncorrelated electrons, which for the Hubbard chain takes the form $\bar{\varepsilon} = -4|t|/\pi$, and $U_c$ is the critical value of $U$, for which we have the Mott-Hubbard transition (i.e. $\langle n_{i\uparrow} n_{i\downarrow} \rangle = 0,$ and $E_G = 0$). This critical value is $U_c = 8|\bar{\varepsilon}|$. Note that at this point the band ($\sim \bar{\varepsilon} < 0$) and Coulomb ($\sim Ud > 0$) contributions to the total system energy compensate exactly each other.

*2.3. Gutzwiller-Wave-Function approximation (GWF)*

The Gutzwiller solution without any further approximation was carried out for the Hubbard chain by Metzner and Vollhardt [7]. In this approach, the ground state energy (2.9) in the $n = 1$ case takes the form

$$\frac{E_G}{N_a} = \varepsilon_a^{eff} - 4|t| \int_{-\pi}^{\pi} dk \cos(k) n_k(g) + Ud(g), \qquad (2.16)$$

where

$$d = \frac{U}{2} \left[ \frac{g}{1-g^2} \right] \left[ \log \frac{1}{g^2} + g^2 - 1 \right], \qquad (2.17)$$

is the number (probability) of double occupancies per site and

$$n_\mathbf{k} = n_\mathbf{k}^0 - (1/2)(1-g)^2 n_\mathbf{k}^0 + \frac{1}{(1+g)^2} \sum_{m=1}^{\infty} (g^2 - 1)^m \left[ 1 - (1-g^2) n_\mathbf{k}^0 \right] f_m(\mathbf{k}) \qquad (2.18)$$

is the statistical distribution function in reciprocal (**k**) space, determined iteratively. In the last expression $n_\mathbf{k}^0 = \theta(k_F - |\mathbf{k}|)$ is the Fermi (Heaviside) function for noninteracting particles at temperature $T = 0$, and the correlation-induced part is given by

$$f_m(k) = \begin{cases} R_m(k) & \text{for } k \leq \pi/4 \\ (-1)^m / 2m + Q_m(k) + Q_m(\pi - k) & \text{for } k > \pi/4 \end{cases} \qquad (2.19)$$

In the above expressions, the coefficients $R_m(k)$ and $Q_m(k)$ can be expressed in terms of the Taylor expansion around the values $k = \pi/4$ and $k = 3\pi/4$ respectively, i.e.

and
$$R_m(k) = \sum_{j=0}^{m} \frac{R_m^{(j)}(\pi/4)}{j!} \left( k - \frac{\pi}{4} \right)^j,$$

$$Q_m(k) = \sum_{j=0}^{m} \frac{Q_m^{(j)}(3\pi/4)}{j!} \left( k - \frac{3\pi}{4} \right)^j. \qquad (2.20)$$

The coefficients $R_m^{(j)}(\pi/4)$ and $Q_m^{(j)}(3\pi/4)$ above can be computed with the help of interactive procedure. Namely, we compute first

$$R_1^{(j)}(\pi/4) = -(1/2)\delta_{j0},$$

$$R_m^{(0)}(\pi/4) = (-1)^m \frac{(2m-1)!!}{(2m)!!}, \quad m \geq 1,$$

$$Q_m^0(\pi/4) = -(-1)^m/(2m) + (-1)^m \frac{(2m-3)!!}{(2m)!!}, \quad m \geq 1, \quad (2.21)$$

$$R_m^{(1)}(\pi/4) = \left((-1)^m/2\right)\frac{(2m-3)!!}{(2m-4)!!}, \quad m \geq 2,$$

$$Q_m^{(1)}(3\pi/4) = \left((-1)^m/2\right)\frac{(2m-5)!!}{(2m-4)!!}, \quad m \geq 2,$$

and then

$$R_{m+1}^{(j)}(\pi/4) = -\frac{m-j+1/2}{m-j+1}R_m^{(j)}(\pi/4) - \frac{1}{2(m-j+1)}Q_{m+1}^{(j+1)}(3\pi/4), \quad 1 \leq j \leq m,$$

$$R_{m+1}^{(m+1)}(\pi/4) = \begin{cases} 0 & \text{for } m+1 \text{ odd} \\ -2Q_{m+1}^{(m+1)}(3\pi/4) & \text{for } m+1 \text{ even}, \end{cases} \quad (2.22)$$

$$Q_{m+1}^{(j+2)}(3\pi/4) = 2(m-2)Q_{m+1}^{(j+1)}(3\pi/4) + 2mR_m^{(j)}(\pi/4)$$
$$- 4j(m-j+1)\left(R_m^{(j)}(\pi/4) + R_{m+1}^{(j)}(\pi/4)\right), \quad 1 \leq j \leq m-1.$$

*2.4. EDABI method*

In the expressions (2.13), (2.15) and (2.16), describing the ground state energy of the Hubbard chain in the half-filled situation, $n = 1$, the quantities $U$ and $t$ are, as a rule, treated as model free parameters and the physical properties of the system are analyzed as a function of both $U/t$ and $n$. This is somewhat problematic, since the parameter $U/t$ is not directly measurable. What is more important, this parameter changes in a nonlinear manner with the increasing interatomic distance, as discussed bellow. In our method of optimized single-particle wave functions, we minimize additionally the system energy with respect to the (restricted) choice of the single-particle wave-functions $\{w_i(\mathbf{r})\}$ contained in the expressions for the parameters $t$ and $U$ defined earlier. So, the single-particle wave function of the Wannier type are *adjusted a posteriori*, in the correlated state, as the electron correlations and the single-particle aspect of the problem are treated on the same footing. This method is suited particularly for correlated nanoscopic systems [3, 12]. In general, it is suited (as it is the case here) for the situations when the interaction is of nonperturbative nature and the expression for the ground-state energy obtained in either analytic form or iteratively. In effect, we can determine the system evolution as a function of interatomic distance.

In this method the Wannier function is composed of atomic 1s functions (the Hubbard model is an effective *s*-band model), which and in the tight-binding approximation takes the form

$$w_i(\mathbf{r}) \equiv \beta\Psi_i(\mathbf{r}) - \gamma\left(\Psi_{i-1}(\mathbf{r}) + \Psi_{i+1}(\mathbf{r})\right), \quad (2.23)$$

where

$$\Psi_i(\mathbf{r}) \equiv \sqrt{\alpha^3/\pi}\exp(-\alpha|\mathbf{r}-\mathbf{R}_i|) \quad (2.24)$$

is the atomic s-wave function with an adjustable $\alpha$ (the Slater s-type orbital). The parameter $\alpha$, which expresses the inverse orbital size ($a \equiv 1/\alpha$), plays the role of the variational parameter in this very simple situation. If we use as $a = a_0$, where $a_0$ is the Bohr radius, the ground state energy is not a minimum! Therefore, the renormalization induced by the electronic correlations is not only possible, but also indispensable. In definition (2.23) the coefficients $\beta$ and $\gamma$ are determined for the translationally invariant systems, i.e. all nearest neighbors are regarded as equivalent. For $R_{ij} \to \infty$ we have than

and
$$\lim_{R_{ij} \to \infty} \alpha = \alpha_0 = 1/a_0,$$
$$\lim_{R_{ij} \to \infty} \beta = 1, \qquad \lim_{R_{ij} \to \infty} \gamma = 0. \tag{2.25}$$

The orthonormalization equations determining $\beta$ and $\gamma$ are

$$<w_i | w_i> = 1 \quad \text{i} \quad <w_i | w_j> = 0, \tag{2.26}$$

and take the form of the system of parabolic equation as $\beta^2 + \gamma^2 = 1$. From four possible solutions we select the following

$$\beta = \frac{A + \sqrt{A^2 - BS_1}}{[2A^2 - BS_1 - zAS_1^2 + 2(A - zS_1^2)\sqrt{A^2 - BS_1}\,]^{1/2}}, \tag{2.27}$$

and

$$\gamma = \frac{S_1}{[2A^2 - BS_1 - zAS_1^2 + 2(A - zS_1^2)\sqrt{A^2 - BS_1}\,]^{1/2}}, \tag{2.28}$$

where
$$A \equiv 1 + S_2, \qquad B \equiv 3 + S_3, \tag{2.29}$$
and
$$S_k \equiv <\Psi_{\mathbf{R}_i} | \Psi_{\mathbf{R}_{i+k}}>. \tag{2.30}$$

is the overlap integral $\langle \Psi_i | \Psi_{i+k} \rangle$. Because of the relation (2.23) the interaction parameters $U$ and $K$ contain combination of four wave functions of type (2.24) centered on different sites. Hence, to make the problem tractable, we approximate the wave functions as the combinations of seven Gaussians (STO-7G basis) and compared them with those for STO-3G basis (combination of the three Gaussians) [12, 13]. The used STO-7G basis did not lead to a qualitative improvement of results when compared to those with STO-3G basis. In Fig. 1 we have shown exemplary optimized Wannier functions centered on nearest neighboring sites and approximated by the Gaussians (STO-3G basis Fig. 1a, and STO-7G basis Fig. 1b).
The parameters $\varepsilon_a$, $t$, $U$, and $K_{ij}$ are as before

$$\varepsilon_a = <w_i | H_1 | w_i>, \qquad t_{i,i+1} = <w_i | H_1 | w_{i+1}> \equiv \int d^3\mathbf{r}\, w_i(\mathbf{r}) H_1 w_i(\mathbf{r}), \tag{2.31}$$

and
$$U = V_{iiii} \equiv <w_i w_i | V | w_i w_i>, \qquad K_{ij} = V_{ijij} \equiv <w_i w_j | V | w_i w_j>, \tag{2.32}$$

where

$$H_1(\mathbf{r}) = -\frac{\hbar^2}{2m}\nabla^2 - \sum_j \frac{e^2}{|\mathbf{r}-\mathbf{R}_j|} \stackrel{a.u.}{=} -\nabla^2 - \sum_j \frac{2}{|\mathbf{r}-\mathbf{R}_j|} \equiv T(\mathbf{r}) - eV(\mathbf{r}). \qquad (2.33)$$

The summation $\sum(...)$ in (2.33) over all lattice sites is restricted to the summation over finite number of sites. The analysis of the site selection in this sum has been discussed by Rycerz [12]. It turns out, that the most effective is the choice: we establish first the coordination sphere $S_k(i)$ of the central site $i$ and that of its nearest neighbor, $S_k(i+1)$. In effect, we can write down the summation in the form

$$H_1(\mathbf{r}) \stackrel{a.u.}{\approx} -\nabla^2 - \sum_{j \in S_k(i) \cup S_k(i+1)} \frac{2}{|\mathbf{r}-\mathbf{R}_j|}. \qquad (2.34)$$

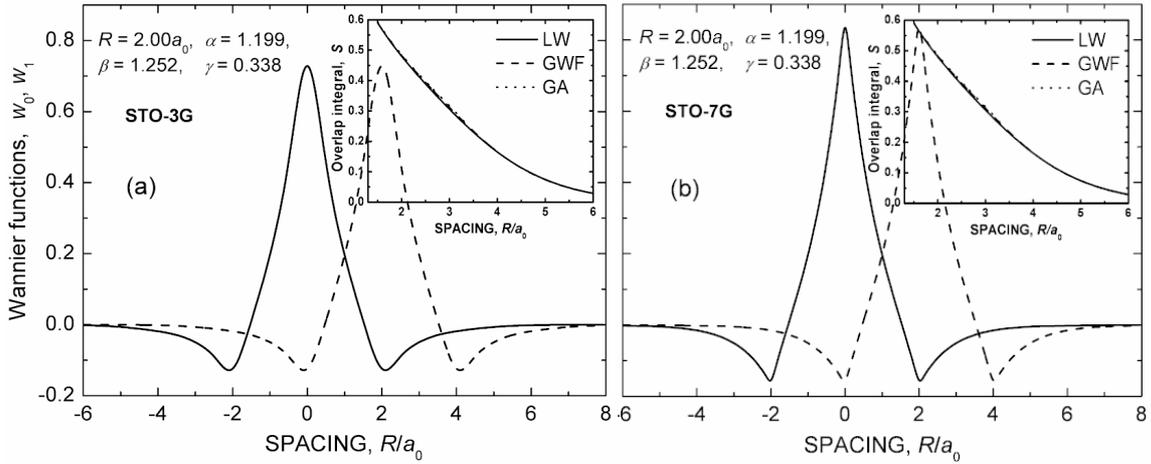

Fig. 1. Two optimized Wannier functions centered on neighboring atoms (solid and dashed lines, respectively), within the STO-3G (a) and STO-7G (b) bases. In the latter case they are more like the Slater orbitals (spikes at the origin and at the minima). Insets: overlap integral vs $R$.; note its value does not depend much on the choice of the method of solution.

Note that number of atomic potentials taken from the right of a given site $i$ is by one larger than that of the potentials taken from the left side of $i$. For 1D-systems, good results are obtained already for $k = 2$, i.e. when we take six potential wells The effective "periodic" potential seen by electron located on site $i$ is depicted in Fig. 2. In this paper we have taken $k = 10$, i.e. ten Coulomb wells from the left and 11 potential wells from the right. We have also made (for comparison) the calculations for $k = 2$ and have compared the results with the previous calculations for nanochains [12, 13]. Our results show that taking 6 Coulomb potential wells is indeed sufficient. This means, that the extension to 22 Coulomb wells did not improve remarkably the accuracy, although it is not essentially difficult (it may be needed for higher than 1s Slater states). An important numerically change may take place only if we extend the superposition (2.23) beyond the nearest neighbors.

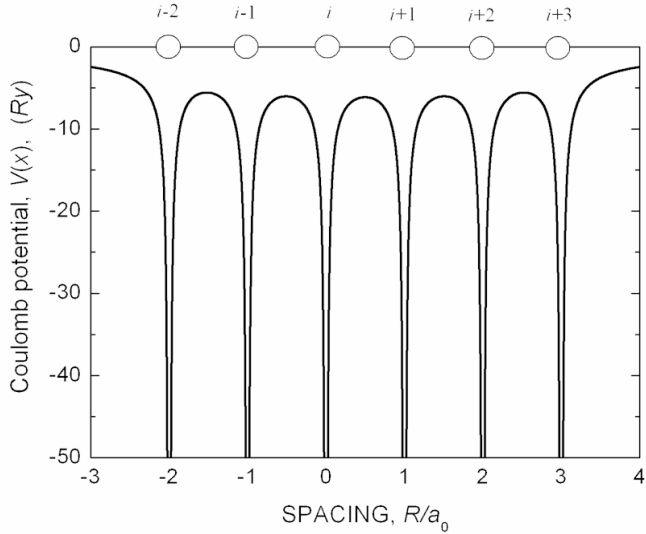

Fig. 2. Effective "periodic" Coulomb attractive potential expressing the construction $S_2(i) \cup S_2(i+1)$ set of sites (see the main text). Note the asymmetry left-right with respect to the site $i$.

In the basis (2.23), involving only the nearest-neighbor overlap [14], the single-particle parameters take the form

$$\varepsilon_a = \beta^2 T_0 - 4\beta\gamma T_1 + 2\gamma^2(T_0 + T_2), \qquad (2.35)$$

$$t \equiv t_{i,i+1} = \beta^2 T_1 - 2\beta\gamma(T_0 + T_2) + \gamma^2(3T_1 + T_3), \qquad (2.36)$$

where

$$T_k \equiv <\Psi_i | H_1 | \Psi_{i+k}>, \qquad (2.37)$$

is the $k$-th hopping integral in the atomic basis.

## 3. Results and discussion

### 3.1. Ground state energy and microscopic parameters

In Fig. 3 we have displayed the ground-state energy (per atom) for the Hubbard chain as a function of intersite distance $R_{ij}$ for the three solutions discussed in the foregoing Section. The system energy is always higher then the energy of isolated hydrogen atoms, as $E_G/N \to -1Ry$ only in the $R \to \infty$ limit. In result, the Hubbard chain is in a vacuum not stable for any finite spacing $R$. Both the Gutzwiller-wave-function and Lieb-Wu solutions converge in the limit $R \to \infty$, although they merge rather slowly with the increasing $R$. The asymptotic, atomic regime is reached approximately for $R > 5.5a_0$, corresponding to $U/|t| \sim 50$ (see the inset to Fig. 3). The function $E_G(R)$ for GA ends up at the interatomic distance $R \approx 3.3a_0 \approx 1.7$ Å (for the optimized 1s Wannier functions). The knowledge of such critical interatomic distance is important for determination of the metallicity of quantum wires. Inset to Fig. 3. shows the nonlinear character of $U/|t|$ vs. $R$. This means that the $U/|t|$ and $R$ dependences of physical quantities is not equivalent. It is important to note here that the calculated $U/|t|$ values within the three methods practically coincide. This remark concerns not only $U/|t|$ values, but also the parameters $U, t, \varepsilon_a^{eff}$ are close, as illustrated in Table 1.

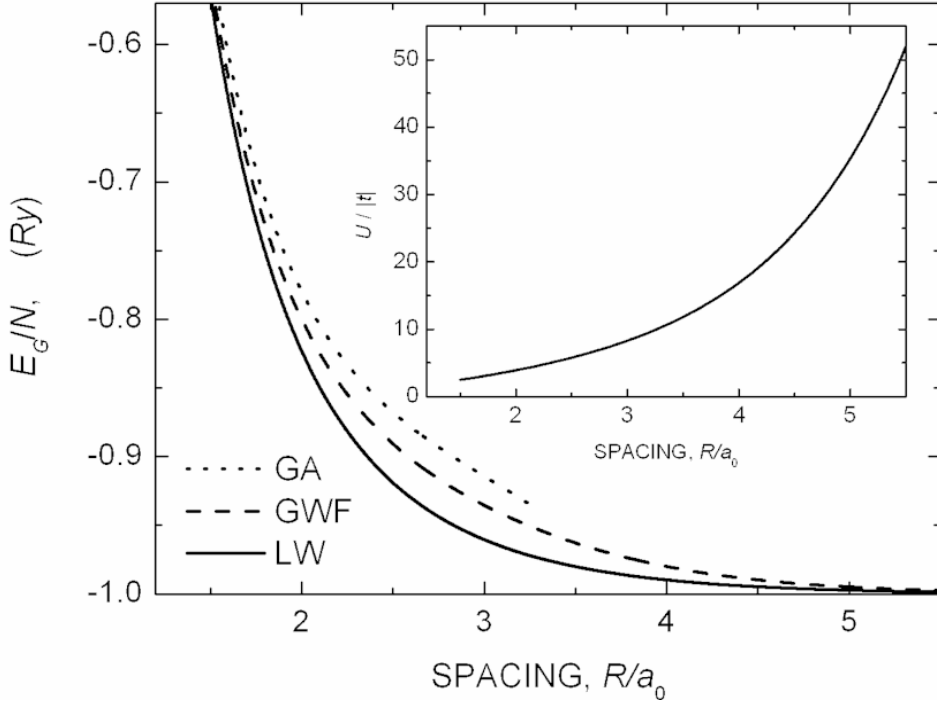

Fig. 3. Ground state energy as a function of *interatomic distance R* for the three methods of solving the Hubbard chain discussed in the main text, with optimization of the single-particle wave-functions. Inset: ratio $U/|t|$ vs. $R$ ; note that $U/|t| \sim 1$ already for $R/a_0 = 1$.

Table 1. Microscopic parameters for the three methods specified in Sec. 2, computed as a function of interatomic distance $R/a_0$

| $R/a_0$ | $\varepsilon_a^{eff}$ (Ry) | | | $t$ (Ry) | | | $U$ (Ry) | | |
|---|---|---|---|---|---|---|---|---|---|
| | LW | GWF | GA | LW | GWF | GA | LW | GWF | GA |
| 1.5 | 0.055 | 0.054 | 0.053 | -0.814 | -0.813 | -0.811 | 2.033 | 2.031 | 2.029 |
| 2.0 | -0.568 | -0.569 | -0.570 | -0.438 | -0.437 | -0.435 | 1.712 | 1.708 | 1.703 |
| 2.5 | -0.804 | -0.805 | -0.806 | -0.265 | -0.262 | -0.261 | 1.527 | 1.517 | 1.510 |
| 3.0 | -0.906 | -0.907 | -0.907 | -0.171 | -0.169 | -0.168 | 1.416 | 1.404 | 1.394 |
| 3.5 | -0.954 | -0.954 | -0.938[a] | -0.114 | -0.114 | -0.133[a] | 1.348 | 1.341 | 1.353[a] |
| 4.0 | -0.977 | -0.977 | | -0.078 | -0.077 | | 1.308 | 1.305 | |
| 5.0 | -0.994 | -0.994 | | -0.036 | -0.036 | | 1.268 | 1.269 | |
| 6.0 | -0.999 | -0.999 | | -0.016 | -0.016 | | 1.255 | 1.255 | |
| 7.0 | -1.000 | -1.000 | | -0.007 | -0.007 | | 1.251 | 1.251 | |
| 8.0 | -1.000 | -1.000 | | -0.003 | -0.003 | | 1.250 | 1.250 | |

[a] for $(R/a_0)_c = 3.288$ metal-insulator transition takes place.

In spite of good agreement displayed in Table 1, the ground-state-energy values differ remarkably, as shown in Fig. 3. The reason of this discrepancy is caused by the behavior of double-occupancy probability $d = \langle n_{i\uparrow} n_{i\downarrow} \rangle$. An illustration of this fact is exemplified in Fig. 4.

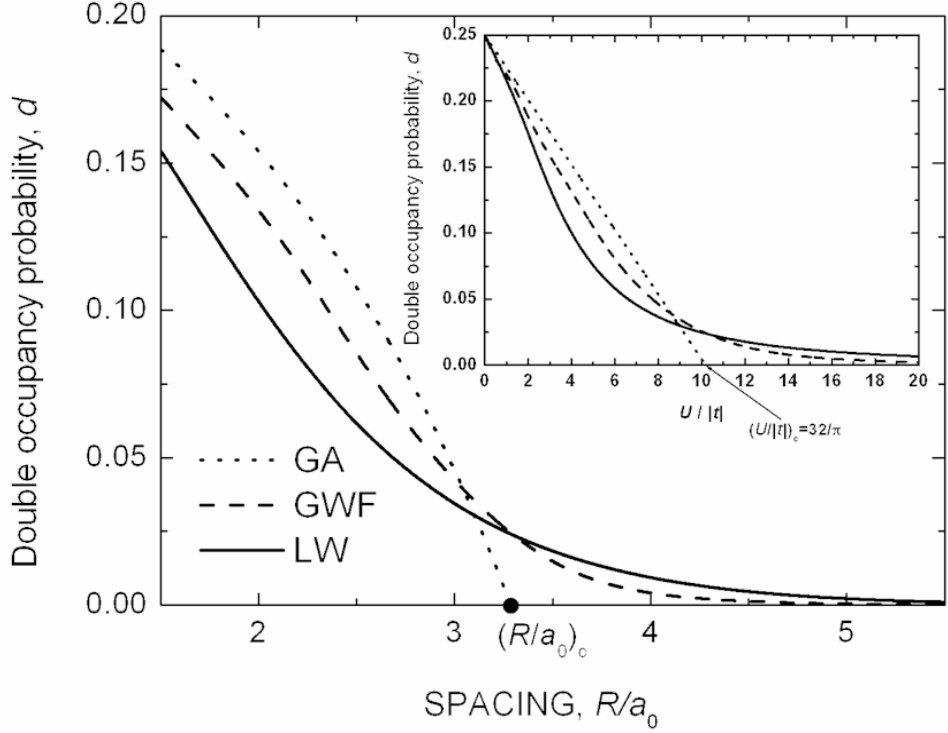

Fig. 4. Double occupancy probability $d = \langle n_{i\uparrow} n_{i\downarrow} \rangle$ vs. interatomic distance for the three methods discussed, with optimized single-particle wave function within STO-7G approach. Inset: double occupancy probability vs. $U/|t|$ calculated for the optimized basis set.

Both the critical values $(R/a_0)_c$ and the $(U/t)_c$ (see the inset) for the metal-insulator transition within GA have also been marked there. In Fig. 5 we show the dependence $E_G/Nt$ vs. $U/|t|$ for the optimized case. The dependence is *essentially linear* in the whole range plotted with the same slope (cf. also the upper inset). For comparison, the dependence of $(E_G - \varepsilon_a^{eff})/N_a|t|$ (the lower inset) versus $U/|t|$ is included (analogical to the Metzner-Vollhardt plot [7]).

We have calculated also the statistical distribution $n_k$ as it evolves with the increasing $R/a_0$, as illustrated in Fig. 6 for selected $R/a_0$ values, i.e. for an experimentally controlled parameter $R$. Note that for $R/a_0 \to 1$, $n_k \to 1$ for $k < k_F = \pi/(2R)$, i.e. we recover the *asymptotic freedom* (Fermi distribution in the high-density limit), and for $R/a_0 \to \infty$, $n_k \to 1/2$ meaning that we have single occupancy with a *definite* spin direction.

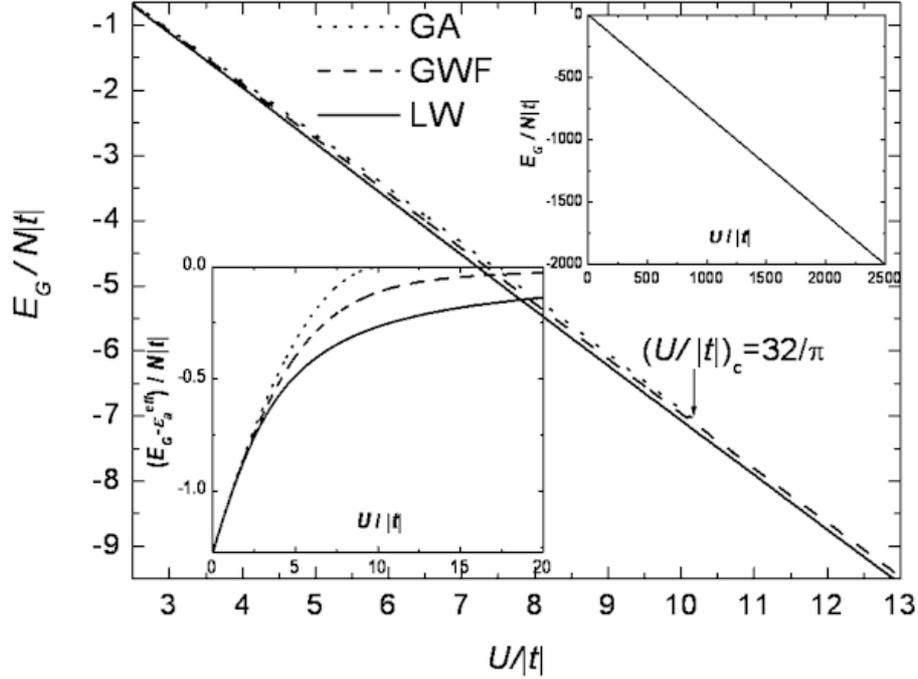

Fig. 5. Ground state energy $E_G/Nt$ vs. $U/|t|$; note the linearity for *all* the calculation method. Upper inset: the linearity of $E_G/Nt$ vs. $U/|t|$ in an extended range (within LW and GWF solutions). Lower inset: The system energy with the subtracted atomic-energy values vs. $U/|t|$ for optimized orbitals.

Finally, we have plotted in Fig. 7 the spacing dependence of the Gutzwiller parameter $g$; the localization regime begins for $R > 4a_0$. In the inset we have shown the corresponding dependence vs. optimized $U/t$ in the same regime.

In GA, the ground state energy is expressed as

$$E_G / N_a = q(d)\bar{\varepsilon} + Ud . \tag{3.1}$$

The first term expresses the renormalized band energy, and the second the electrostatic repulsion energy. $\bar{\varepsilon}$ represents the kinetic (band) energy per atom for noninteracting electrons, whereas the quantity $q$ is the so-called *band narrowing factor* or many-body quasiparticle-mass renormalization. The narrowing factor is in the interval $0 \le q \le 1$; in GA it takes the form when $d \ge 0$ (i.e. in the metallic phase):

$$q(d) = 8d(1-2d) = 1 - (U/U_c)^2 . \tag{3.2}$$

The value $U = U_c$ determines the so-called *Brinkman-Rice point* [15, 16]. In GWF we have to calculate either $q(U/U_c)$ or $q(R)$ numerically, as the method does not lead to the analytical results in the closed form. On other hand, LW original solution can be expressed in the form

$$\frac{E_G}{N_a} = -|\bar{\varepsilon}|q(d) + Ud = -4|t|\int_0^\infty d\omega \frac{J_0(\omega)J_1(\omega)}{\omega(1+\exp(\omega U/(4|t|)))} . \tag{3.3}$$

The double occupancy probability is also determined analytically and is

$$d \equiv \frac{1}{N_a} \frac{\partial E_G}{\partial U} = \int_0^\infty d\omega \frac{J_0(\omega) J_1(\omega)}{1 + \cosh(\omega U/(2|t|))} \tag{3.4}$$

and $\bar{\varepsilon} = -4|t|/\pi$. Finally, we obtain the expression for the band narrowing

$$q = \pi \int_0^\infty d\omega \frac{J_0(\omega) J_1(\omega)}{\omega(1 + \exp(\omega U/(4|t|)))} + \frac{\pi U}{4t} \int_0^\infty d\omega \frac{J_0(\omega) J_1(\omega)}{1 + \cosh(\omega U/(2|t|))}. \tag{3.5}$$

The $q(d)$ dependence is displayed in Fig. 8.

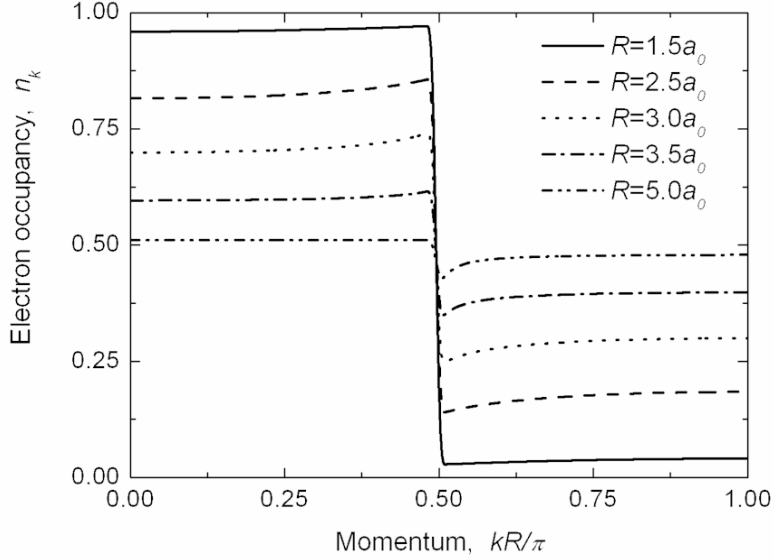

Fig. 6. Momentum distribution $n_k$ vs. $k$ for different lattice parameter within GWF approximation.

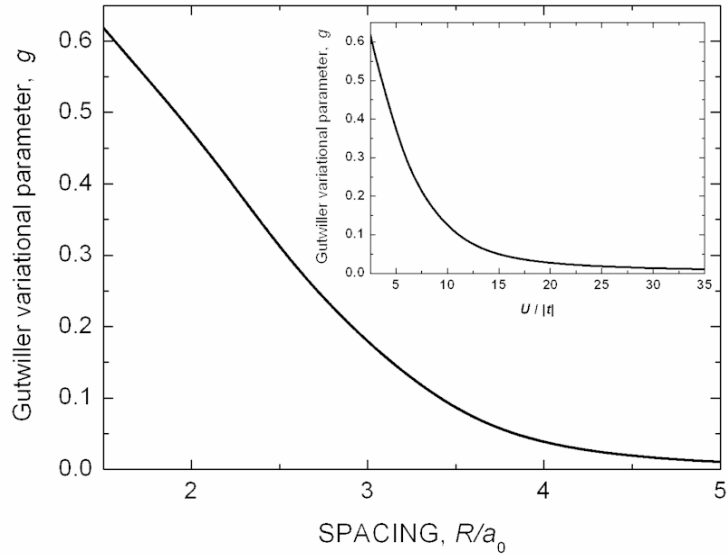

Fig. 7. Gutzwiller variational parameter vs. interatomic distance; the corresponding dependence vs. $U/|t|$ is qualitatively similar, as demonstrated in the inset.

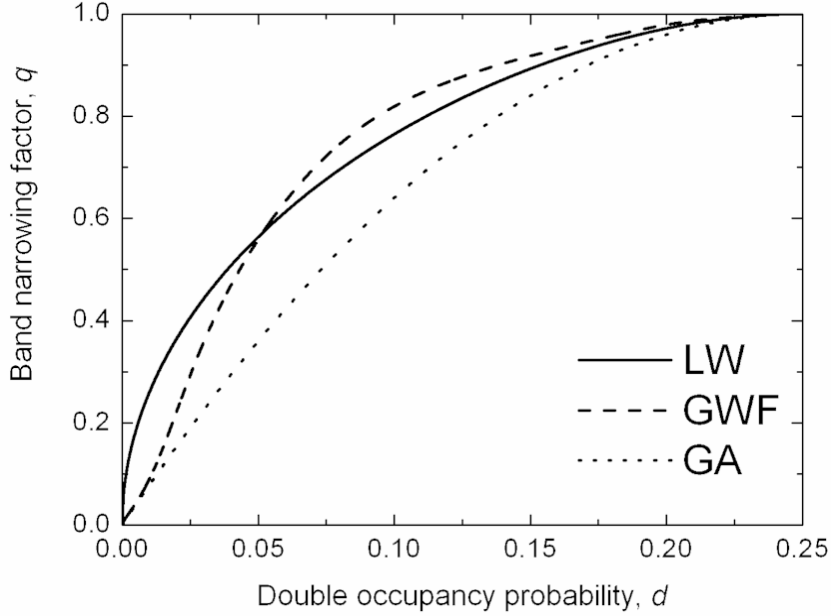

Fig. 8. Double occupancy dependence of band narrowing factor within the three schemes discussed in main text.

*3.2. Comparison between Lieb-Wu and EDABI solutions for nanochain*

We have compared the LW results with those from EDABI method for periodic boundary and observe quite good agreement between them provided the same boundary conditions (periodic) are considered. The actual numbers listed in Table 2 are for the chain of $N = 10$ atoms containing one electron (of 1s type) per site. The computed quantities are obtained starting from STO-3G basis and taking into account six Coulomb potential wells.

**Concluding remarks**

In this paper we have included the single-particle wave function optimization on the same footing as the diagonalization of the many-body Hamiltonian in the Fock space. The correlation-included renormalization of the wave function is essential, since the effective Bohr radius of atomic functions, composing the Wannier function, can be reduced by about 30% (cf. Table 2 for the values of $1/\alpha_{min}$). We can say that our method *completes* the solution of solvable many-body Hamiltonians, as the model parameters are determined as a function of interatomic distance (cf. Table 1). Other models such as Anderson-impurity model, for which an exact solution is available [17] can be treated in the same manner, as the Lieb-Wu solution for the Hubbard chain here. A separate branch of research is concerns application of EDABI method to correlated nanosystems [3].

Table 2. Selected parameters of the Hubbard Hamiltonian and ground state energy for LW solution ($N = \infty$) for both STO-3G or STO-7G single-particle variational basis, as well as these for a nano-chain containing $N = 10$ atoms.

| $R/a_0$ | $\alpha_{min}$ (a.u.) | | | $\varepsilon_a^{eff}$ (Ry) | | | $E_G$ (Ry) | | |
|---|---|---|---|---|---|---|---|---|---|
| | $N = \infty$, 7G | $N = \infty$, 3G | $N = 10$, 3G | $N = \infty$, 7G | $N = \infty$, 3G | $N = 10$, 3G | $N = \infty$, 7G | $N = \infty$, 3G | $N = 10$, 3G |
| 1.5 | 1.332 | 1.306 | 1.309 | 0.055 | 0.130 | 0.131 | -0.565 | -0.553 | -0.568 |
| 2.0 | 1.182 | 1.204 | 1.205 | -0.568 | -0.535 | -0.534 | -0.824 | -0.810 | -0.815 |
| 2.5 | 1.100 | 1.120 | 1.120 | -0.804 | -0.789 | -0.789 | -0.919 | -0.912 | -0.914 |
| 3.0 | 1.055 | 1.068 | 1.067 | -0.906 | -0.897 | -0.898 | -0.960 | -0.956 | -0.957 |
| 4.0 | 1.018 | 1.021 | 1.020 | -0.977 | -0.970 | -0.970 | -0.990 | -0.984 | -0.984 |
| 5.0 | 1.006 | 1.006 | 1.005 | -0.994 | -0.987 | -0.987 | -0.997 | -0.990 | -0.990 |
| 6.0 | 1.002 | 1.002 | 1.003 | -0.999 | -0.991 | -0.992 | -0.999 | -0.991 | -0.991 |
| 7.0 | 1.000 | 1.001 | 1.000 | -1.000 | -0.992 | -0.992 | -1.000 | -0.992 | -0.992 |
| 8.0 | 1.000 | 1.000 | 1.000 | -1.000 | -0.992 | -0.992 | -1.000 | -0.992 | -0.992 |

**Acknowledgment**

We thank Adam Rycerz for discussion on the Gaussian basis. The work was supported by the Ministry of Higher Education and Science, Grant No. 1P03B 001 29. One of the authors (J.S.) was supported also by Foundation for Polish Science (FNP). This work represents a part of the Ph.D. Thesis of one of the authors (J.K.) to be submitted to the Faculty of Physics, Astronomy, and Applied Computer Science of the Jagiellonian University.